\documentstyle[12pt]{article}
\input math_macros
\begin{document}

\begin{titlepage}
\begin{center}
August 1992
\hfill       LBL-32389 \\
(Revised)                    \hfill          UCB-PTH-92/19 \\
\vskip 1.0 in
{\large \bf INTERPRETATION OF PRECISION MEASUREMENTS IN THE STRONGLY
INTERACTING
LIMIT OF THE STANDARD ELECTROWEAK MODEL }\footnote{This work was supported by
the Director, Office of Energy Research, Office of High Energy and Nuclear
Physics, Division of High Energy Physics of the U.S. Department of Energy
under contract DE-AC03-76SF00098 and in part by the National Science
Foundation under grant PHY--90--21139.}\\
\vskip .5 in
{
{\bf Mary K. Gaillard}\\
\vskip 0.5 cm

{\it Department of Physics, University of California\\ and\\
Physics Division, Lawrence Berkeley Laboratory, 1 Cyclotron Road\\
Berkeley, CA 94720 \\ } }
\end{center} \vskip 1.0  in \begin{abstract}
Strong rescattering corrections to one-loop contributions to the parameters of
the standard electroweak model are considered.
\end{abstract}
\end{titlepage}

\newpage

\def\M{{\cal M}}
\def\L{{\cal L}}
\renewcommand{\thepage}{\arabic{page}}
\setcounter{page}{1}
\setcounter{equation}{0}

Once the top quark mass is known, precision electroweak measurements
can be used to constrain the value of the Higgs mass
in the context of the standard model.
The results are generally quoted for Higgs masses up to $1TeV$.
However, for such a large mass the Higgs sector of the theory is strongly
interacting~\cite{tini}, and the one-loop perturbation calculation may not be
very meaningful.  In fact, a Higgs-like state with mass $\gg m_W$ should
be regarded as a resonance in the $I=J=0$ channel of longitudinally
polarized vector meson scattering, rather than as an elementary particle.
More generally, if there is no such scalar resonance, there is a
contribution to quantum corrections to electroweak parameters from a strongly
interacting sector, for which the low energy effective theory is the gauged
nonlinear $\sigma$-model that describes~\cite{appel},~\cite{chan} the
couplings of longitudinally polarized vector bosons.

In this paper I reinterpret the ``Higgs sector'' contribution to radiative
corrections to electroweak parameters in the context of this gauged nonlinear
$\sigma$-model, where the goldstone bosons of the strongly interacting sector
are the longitudinally polarized $W,Z$~\cite{appel},~\cite{chan}.  The one-loop
approximation is replaced by contributions that are of lowest order in the weak
couplings $g$ of transversely polarized bosons, but of arbitrary order in the
strong self-interactions of the effective $\sigma$-model.  This is implemented
by the use of dispersion relations, in which the integrands involve
on-shell $S$-matrix elements; therefore the equivalence theorem~\cite{chan}
allows us to replace the transversely polarized bosons in the intermediate
states by pseudoscalars that are hereafter referred to as pions ($\pi$).

I adopt the point of view that the nonlinear $\sigma$-model is an effective
theory that is valid only below some energy scale $\Lambda$, and the relevant
dispersion integrals are cut off at that scale.  Thus I define the ``Higgs
sector'' contribution as those contributions to the relevant spectral functions
involving $n\pi$ intermediate states in the energy range $m_W^2<s< \Lambda^2$.
In technicolor models, for example, there are important contributions from
technifermion loops at higher energies~\cite{stu}.  More generally, the
theory that emerges above the effective cut-off for the nonlinear sigma-model
may have other degrees of freedom that would contribute to the high energy
parts of the spectral functions, and these would have to be considered
separately in the context of a specific theory.

As will be shown below,
the contributions to the observable parameters $S,T,U$~\cite{stu} arise
only from $J\le 1,\; I\le 2$ multipion intermediate
states.  Contributions from the $n\pi$ channels that have quantum numbers
analogous to the $\rho,a_1,\pi,\sigma$ hadron states
can be inferred from existing one loop perturbative
calculations.  In the Landau gauge, in which the distinction between
(weakly coupled) transverse vector and (strongly coupled) scalar degrees of
freedom is unambiguous, the Feynman integrals that determine the one-loop
``Higgs contribution'' can be isolated and rewritten as
dispersion integrals.  Summation to all orders in strong rescattering in the
lowest partial waves is achieved by replacing the integrands with the
squared matrix elements for $W_T\to n\pi$ and $W_T\to n\pi+W_T$ in the
appropriate $n\pi$ channels.  In the approximation that elastic
scattering saturates strong interaction unitarity over the relevant energy
range, these matrix elements are then determined by low energy constraints
due to the chiral symmetry~\cite{appel},~\cite{chan} of the effective
$\sigma$-model and by the strong scattering phase shifts.

The parity even $I= 2,\;J=0
\;n\pi$ channel may also contribute at subleading order
in the weak coupling constant $g$.  However power counting
can be misleading in a channel with a sufficiently narrow resonance,
as will be illustrated by the analysis of the $\sigma$ channel
(parity even $I=J=0$).

I first recall the one-loop results.
The large $m_H$ ``Higgs contribution'' to $S,T,U$~\cite{stu}
is usually defined by subtracting out the standard model
result evaluated at some reference value $m$ of the Higgs mass,
of the order of $m_W$, that is, the standard model
``Higgs contribution'' (SM) is defined as the
piece which grows with the with Higgs mass; the one-loop result is~\cite{stu}
\begin{equation}
S_{SM}^{(1)} = {1\over 12\pi}\ln\left({m_H^2/m^2}\right), \;\;\;\;
T_{SM}^{(1)} = -{3\over 16\pi cos^2\theta}\ln\left({m_H^2/m^2}\right),
\;\;\;\; U_{SM}^{(1)} = 0.
\end{equation}
In the one-loop approximation, the analogous contribution in the nonlinear
$\sigma$-model formulation is that contribution which grows with the cut-off
$\Lambda$.  The divergent part of the one-loop effective lagrangian for the
$SU(1)\times U(1)$ gauge nonlinear $\sigma$-model (NL)
is given in~\cite{oren},
where the scalar sector is represented by the complex isospinor:
\begin{equation} \Phi = {1\over \sqrt{2}}(\sigma + i\vec\pi\cdot\vec\tau)
\pmatrix{0\cr 1\cr}, \;\;\;\; \sigma = \sqrt{v^2 -\vec\pi^2}.
\end{equation}
Writing the effective tree plus one-loop lagrangian of~\cite{oren} in the
unitary gauge ($\vec\pi = 0$), the relevant part is
\begin{equation}
\L = -{1\over 4}F^a_{\mu\nu}F^{\mu\nu}_b\Pi'_{ab}(0) + {1\over 2}A_\mu^aA^\mu_b
\Pi_{ab}(0) + {\rm higher \;derivative \;terms}.
\end{equation}
The ultraviolet divergent contributions were evaluated in~\cite{oren}; these
give for the observable parameters~\cite{stu} $A^i = S,T,U$, which are linear
combinations of the corrections $\delta\Pi(0)$ and $\delta\Pi'(0)$ to the
inverse vector boson propagators $\Pi_{ab}(s) = \Pi_{ab}(0) +
s\Pi'_{ab}(0) + O(s^2)$,
\begin{equation}
S_{NL}^{(1)} = {1\over 12\pi}\ln\left({\Lambda_S^2/m^2}\right), \;\;\;\;
T_{NL}^{(1)} = -{3\over 16\pi cos^2\theta}\ln\left({\Lambda_T^2/m^2}\right),
\;\;\;\; U_{NL}^{(1)} = 0,
\end{equation}
where $m\approx m_W$ is the effective infrared cut-off, and the ultraviolet
cut-offs $\Lambda_S,\Lambda_T\sim \Lambda$ include an {\it a priori} unknown
correction from the finite contribution.  As previously
noted~\cite{oren},~\cite{long}, the one-loop corrections in the
effective nonlinear $\sigma$-model reproduce those found~\cite{bij} in the
standard model, provided the cut-offs are replaced by the Higgs mass.

In the Landau gauge, $S_{NL}^{(1)}$ and $T_{NL}^{(1)}$ correspond
to Feynman diagrams with $2\pi$ and $W_T+\pi$ intermediate
states, respectively.  The same diagrams give identical
contributions to the standard model result (1), where the ultraviolet
divergence
is cancelled by the $H+\pi$ and $H+W_T$ intermediate state contributions to
$S$ and $T$, respectively.  In the nonlinear model, multipion intermediate
states with the same quantum numbers as these contribute to the same linear
combinations of propagator corrections, so the full contributions
of these states in the nonlinear model, including strong rescattering
corrections, are directly related to the
one-loop results.  The result for these contributions takes the form:
\begin{equation}
U_{NL}= 0,$$
$${g^2\over 16\pi}S_{NL} = {1\over 2\pi}\int_{m^2}^{\Lambda^2_\rho}
 {dt\over t^2}|\M_\rho(t)|^2 - {1\over 2\pi}\int_{m^2}^{\Lambda^2_a}
 {dt\over t^2}|\M_a(t)|^2\equiv {g^2\over 16\pi}(S_\rho + S_a),
\end{equation}
and
\begin{equation}
{cos^2\theta g^2m_W^2\over 4\pi}T_{NL} =
{1\over 2\pi}\int_{m^2}^{\Lambda^2_\sigma}
{dt\over t}|\M_\sigma(t)|^2 - {1\over 2\pi}\int_{m^2}^{\Lambda^2_\pi}
{dt\over t}|\M_\pi(t)|^2$$
$$\equiv {cos^2\theta g^2m_W^2\over 4\pi}(T_\sigma + T_\pi), \end{equation}
where $\Lambda_i \sim\Lambda$ are the effective cut-offs and the matrix
elements $\M_i$ will be defined below.

The derivation of (5) is straightforward, and the result is completely
general.  To lowest order in $g$, transverse
gauge bosons couple to the strongly interacting $\pi$-sector via the $I=1$
vector and axial currents $\vec V_\mu, \vec A_\mu$~\cite{chan}:
\begin{equation}
\L\ni eA^\mu_{(\gamma)}V^3_\mu + {g\over 2}\left[W^{+\mu}(V^-_\mu + A^-_\mu) +
{\rm h.c.}\right] + {g\over 2cos\theta}Z^\mu\left[A^3_\mu +
V^3_\mu(1-2sin^2\theta)\right],\end{equation}
where $eA^\mu_{(\gamma)}$ is the photon field. The currents $\vec V_\mu,
\vec A_\mu$ transform as triplets under the isospin of the
strongly coupled pion sector and their normalization is fixed by low energy
theorems.

$S$ and $U$ are linear combinations of the $\Pi'(0)$'s in (3) which
appear only in the transverse part of the inverse propagator, and therefore are
determined by the $J=1$ spectral functions $<VV>$ and $<AA>$.
Writing the inverse propagator in the form
\begin{equation}
\Pi(p)_{\mu\nu} = g_{\mu\nu}\Pi_0(p^2) + (g_{\mu\nu}p^2-p_\mu p_\nu)\Pi_1(p^2)
= \left(g_{\mu\nu} - {p^\mu p^\nu\over p^2}\right)\Pi(p^2) +
{p^\mu p^\nu\over p^2}\Pi_0(p^2), $$
$$ \Pi(p^2) = \Pi_0(p^2) + p^2\Pi_1(p^2), \;\;\;\; \Pi(0) = \Pi_0(0),
\end{equation}
using the definitions~\cite{stu}~\cite{lang} of $S,U$,
and assuming that the transverse part $\Pi(s)$ satisfies a once-subtracted
dispersion integral gives (5), where
$|\M_i(s)|^2$ is the squared invariant matrix element, integrated over final
state momenta for virtual $W_T$ decay: $W_T(s)\to (n\pi)_i$.  The index
$i$ denotes the quantum numbers of the multipion state, namely the $J=I =1$
parity odd and even state for $i= \rho,A$ respectively.

Since the nonlinear $\sigma$-model is not a renormalizable theory, a
completely general treatment requires an infinite summation over
intermediate states with arbitrary numbers of $\pi$'s and
unknown coupling constants that scale with appropriate powers of an inverse
mass parameter.  Their evaluation would require specification of the full
low energy theory.  Realistically,
the cut-off $\Lambda$ is expected to be less than a few $TeV$.
Low energy theorems dictate that near threshold multipion production costs a
factor of
\begin{equation}
\epsilon =(s/16\pi^2v^2)\approx (\sqrt{s}/3TeV)^2
\end{equation}
for each additional pion.  Therefore few-pion intermediate states are
expected to
be dominant if the effective cut-off $\Lambda$ is below the threshold for
multi-resonance production.  Here I will evaluate these amplitudes under the
assumption that the dominant contributions for $s\le
\Lambda$ are from the fewest-body $n\pi$ states with the appropriate quantum
numbers.  A more general treatment
could be warranted if a specific model or data from the SSC indicated that
multi-pion intermediate states should be significant.

In this approximation $n=2$ for $i=\rho$, and $|\M_\rho(s)|^2 =
g^2s|f_\rho(s)|^2/96\pi$, where $f_\rho(s)$ is the single form factor which,
by virtue of the chiral symmetry of the strongly interacting sector,
satisfies~\cite{chan} $f_\rho(0) =1.$  If only two-body intermediate states
are important, the effects of rescattering can be incorporated using the
Omn\`es equation~\cite{omnes}:
\begin{equation}
|f_\rho(s)| = {\rm exp}\left({s\over \pi}P\int {dt\over t}{\delta_{11}(t)
\over t-s}\right)  ,\end{equation}
where $\delta_{11}$ is the $I=J=1$ scattering phase shift, and $P$ stands for
principal value.  The free-field limit $f_\rho(s) = 1$ gives the
first equation in (2).  For $i=A$, $n=3$, and $\M_a(t)$ is again determined by
a single form factor~\cite{chan}, $f_a(0)$ = 1.
In this approximation the total contribution to $S$ is
\begin{equation}
S_{NL} = {1\over 12\pi}\int_{m^2}^{\Lambda^2} {dt\over t}|f_\rho(t)|^2
-{1\over 256\pi^3}\int_{m^2}^{\Lambda^2} {dt\over t}|f_a(t)|^2,
\end{equation}
with
\begin{equation}
|f_a(s)| = {\rm exp}\left({s\over \pi}P\int {dt\over t}
{\delta_a(t)\over t-s}\right) ,
\end{equation}
where $\delta_a$ is the $I=J=1$ scattering phase shift in the axial-vector
3-body channel.

To obtain the expression (6) for $T$, which is defined
by~\cite{stu},~\cite{lang}
\begin{equation}
\alpha T = {1\over m^2_W}[\Pi_{W^+W^-}(0) - cos^2\theta\Pi_{ZZ}(0)
-2sin\theta cos\theta\Pi_{\gamma Z} -2sin^2\theta\Pi_{\gamma\gamma}],
\end{equation}
note that the result (3) was obtained by an expansion around
non-gauge-fixed background fields; the quantum-corrected effective
Lagrangian obtained in this way
is fully $SU(1)\times U(1)$ gauge invariant.  As a consequence the
two point functions extracted from this Lagrangian satisfy the Ward identities:
\begin{equation}
p_\mu p_\nu\Pi_{\gamma\gamma}^{\mu\nu}(p) =
p_\mu p_\nu\Pi_{\gamma Z}^{\mu\nu}(p) = 0, $$
$$p_\mu p_\nu\Pi_{W^+W^-}^{\mu\nu}(p) =
{1\over 4}g^2v^2\Gamma_{\pi^+\pi^-}(p^2),
\;\;\;\; p_\mu p_\nu\Pi_{ZZ}^{\mu\nu}(p) =
{1\over 4cos^2\theta}g^2v^2\Gamma_{\pi^0\pi^0}(p^2),
\end{equation}
or
\begin{equation}
p^2\Pi^{\gamma\gamma}_0(p^2) = p^2\Pi^{\gamma Z}_0(p^2) = 0, $$
$$p^2\Pi^{W^+W^-}_0(p^2) = {1\over 4}g^2v^2\Gamma_{\pi^+\pi^-}(p^2),
\;\;\;\; p^2\Pi^{ZZ}_0(p^2) =
{1\over 4cos^2\theta}g^2v^2\Gamma_{\pi^0\pi^0}(p^2),
\end{equation}
and since $\Pi(0)= \Pi_0(0)$, $T = T(0)$ where
\begin{equation}
\alpha T(p^2) = {1\over m^2_W}[\Pi_0^{W^+W^-}(p^2) -
cos^2\theta\Pi_0^{ZZ}(p^2)] = $$
$$p^{-2}[\Gamma_{\pi^+\pi^-}(p^2) - \Gamma_{\pi^+\pi^-}(p^2)]\Bigg|_{p^2 \to 0}
. \end{equation}
These Ward identities
hold to all loop order in the strong interactions provided a similar
gauge invariant procedure is followed. The definition (13) assures that the
couplings (7) do not contribute to T in lowest order; the
intermediate states that contribute at lowest order in $g$ are
$W_T + (n\pi)_i.$  This is also evident from the form (16)
of the Ward identity: the chiral symmetry of the strong sector assures that the
right hand side vanishes if the effects of transverse boson exchange are not
included in the pion propagator.  The lowest order contributions to $T$
therefore take the form
$$\Pi(0)={1\over 4}g_{\rho\sigma}\int{d^4p\over p^2-m_W^2}\left(g_{\mu\nu}
- {p_\mu p_\nu\over p^2}\right)\M^{\rho\mu\sigma\nu}(0,p,0,p),$$
where $\M^{\rho\mu\sigma\nu}(0,p,0,p)$ is the forward scattering amplitude for
off-shell vector mesons, {\it e.g.}, $W^\rho_T(0) + W^\mu_T(p)\to
W^\sigma_T(0) + W^\nu_T(p).$  These amplitudes can be expressed as dispersion
integrals where the absorptive part is $\sum_i\M^{\rho\mu i}(0,p)
{\overline{\M}}^{\sigma\nu}_i(0,p)$, {\it i.e.}, the squared amplitude for
$W_T(0)\to (n\pi)_i + W_T(p)$.  It is easy to see that the only nonvanishing
contribution is from a multipion state with $J= 0$:  $\M^{\rho\mu i}(0,p)
= g^{\rho\mu}\sqrt{{4\over 3}}\M_i(p^2)$.  (A factor $3/4$ from the Landau
propagator in the Feynman integral has been absorbed into the definition of
$|\M_i|^2$.)  Moreover, it follows from the Ward
identity (16) that the absorptive part of $\Pi(0)$ contains the same spectrum
as the absorptive part of $\Pi_\pi$, {\i.e.}, only those states with with the
$C$ and $I$ quantum numbers allowed in lowest order by the couplings (7) for
the transition $\pi\to W_T + (n\pi)_i$.  Restricting these to $I= 0,
\; P = 1$ and $I= 1,\; P= -1 \;(i= \sigma,\pi,$ respectively), gives (6).

Generalization of this result to include the only other allowed state, namely
$I = 2,\;P=1,$ is straightforward, However,
the only order $g^2$ contribution to $T$ is in fact
from $(n\pi)_\pi$.  Although $|\M_\sigma|^2$ is formally of order $g^4$, the
standard model Higgs contribution (1)
to $T$ should be recovered in the limit where $\pi\pi$ scattering in the
$I=J=0$ channel is dominated by by a scalar resonance with the width-mass
relation of the standard model Higgs particle.
I therefore include the scalar channel in
(6) in order to encompass this possibility.  Then,
provided the $I=2$ channel has no similar
resonance below the effective cut-off, the
result (6) is again completely general.

Here I will evaluate the integrals
using the same ``few body'' approximation as for $S$.
The contribution (4) to $T$ from the $\pi+W_T$
intermediate state has no
strong rescattering corrections. The first strong correction arises from
the $3\pi + W_T$ pseudoscalar intermediate state, which is probably a smaller
effect than the $3\pi$ axial vector contribution to $S$; I will neglect it
here.  It can be incorporated using the same procedure as for the
$2\pi + W_T$ state to be considered below.

In the standard model,
the Higgs contribution to $T$ comes from a Higgs +
vector loop that arrises from the coupling $(g^2/4)(\sigma^2+\vec\pi^2)
(|W^2| + Z^2/2cos^2\theta)$. In the linear theory $H = \sigma -v$, so there is
a coupling $(g^2/2)H(|W^2| + Z^2/2cos^2\theta)$. In the nonlinear limit (1),
this reduces to the mass term, and there is no contribution to $T$ from this
term.  This is equivalent to integrating out the Higgs field $H$ in the linear
theory;  the point coupling $\vec\pi^2(|W^2| + Z^2/2cos^2\theta)$ is cancelled
by the Higgs exchange contribution to the same coupling for $m_H^2\gg u$,
where $u$ is the invariant $\pi\pi$ squared mass.
In a general nonlinear $\sigma$-model one has to include higher dimensional
operators~\cite{appel},~\cite{long}.  For example, the operator
$M^{-2}|D^2 \Phi|^2$ induces a term
\begin{equation}
\L\ni {g^2\over 4}M^{-2}
\partial^\mu\vec\pi\cdot\partial_\mu\vec\pi(|W^2| + Z^2/2cos^2\theta)
\end{equation}
which gives a contribution $|\M_\sigma(t)|^2 =
\int(9g^2m_W^2u^2/8v^2M^4)|f_\sigma(u)|^2LIPS $, where I normalize the
form factor for $W\to W_T + (2\pi)_\sigma$ by $f(u)$ = 1.  The dimensional
coefficient in (17) has been chosen to match that which
would arise from integrating out a standard model Higgs particle with mass
$m_h = M$, but $M$ is in general arbitrary.  I will assume in
what follows that this is the leading contribution to $T_\sigma$.
Writing the Lorentz invariant phase space integral as
\begin{equation}
LIPS = \int\prod_{i=1}^3\left({1\over (2\pi)^3}{d^3p_i\over 2E_i}\right)
(2\pi)^4\delta^3(\sum p^i)\delta(E_i - \sqrt{t}) = {1\over 32\pi^2}\int_{m^2}^t
{du\over 2\pi}{t-u\over 2t},
\end{equation}
inserting this in (6) and performing the $t$-integration, we get in the
``fewest body'' approximation:
$$ T_{NL} = T_\sigma -
{3\over 16\pi cos^2\theta}\ln\left({\Lambda_T^2/m^2}\right), $$
$$T_\sigma =
{9\over 512\pi^3v^2M^4cos^2\theta}\int_{m^2}^{\Lambda^2}dt|f_\sigma(t)|^2
t^2[\ln(\Lambda^2/t) - {t\over \Lambda^2}+ 1],$$
\begin{equation}
|f_\sigma(s)| = {\rm exp}\left({s\over\pi}P\int{dt\over t}{\delta_{00}(t)\over
t-s}\right) ,\end{equation}
where $\delta_{00}$ is the $I=J=0$ scattering phase shift.

Note that the Omn\`es equation used here was derived~\cite{omnes} for form
factors assumed to satisfy once-subtracted dispersion relations. In general
the form factors written here could be multiplied by polynomials
$f_i(s)\to P_i(s)f_i(s),\;P_i(0)= 1$,
with the degree of $P_i$ determined by the number of subtractions
required~\cite{omnes}.  In the present context, since all integrals are
cut-off at $s\sim\Lambda^2$ the question of ultraviolet convergence is
irrelevant, but higher order polynomials in $s/M^2$ could correspond to the
presence of additional operators of higher dimension.

I now consider, as illustrative examples, two extreme limits of the phase
shifts:  extrapolation of the low energy theorems and resonance dominance.
A convenient (and entirely general) parametrization
of the phase shifts is obtained by writing the partial wave amplitudes as
\begin{equation}
a_i(s) = -{\lambda_i(s)\over s - M_i^2(s) +i\lambda_i(s)}, \;\;\;\;
\delta_i(s) = {1\over 2i}\ln\left({s - M_i^2(s)
-i\lambda_i(s)\over s - M_i^2(s) + i\lambda_i(s)}\right)
\end{equation}
This satisfies unitarity: $a_i(s) = sin\delta_i(s)e^{i\delta_i(s)}$,
and the behavior of $\lambda_i/M_i^2$
for $s\to 0$ is determined by low energy theorems~\cite{appel},~\cite{chan}:
$$ a_\sigma \equiv a_{00} =
\pi\epsilon\left(1 +{25\over 18}\epsilon\ln(\Lambda^2/s)
+ i\pi\epsilon\right)[1+ O(\epsilon)] $$
$$ a_\rho \equiv a_{11} = {\pi\epsilon\over 6}\left(1 + i{\pi\epsilon\over 6}
\right)[1+ O(\epsilon)] $$
\begin{equation}
a_a \sim {\pi\epsilon^2\over 8},
\end{equation}
where $\epsilon$ is defined in (9), and the $O(s^2)$
terms in $a_\sigma$ and $a_\rho$ are the one-loop corrections~\cite{mko} to the
effective low energy theory.

If $s-M^2_i(s)$ does not vanish for $s
<\Lambda^2$, there is no resonance and the amplitudes (12) are just
unitarizations of the amplitudes extrapolated from low energy theorems.
The calculation of $S,T,U$ using the effective chiral lagrangian has been
considered in~\cite{maria}.
If we retain only the leading order in $\epsilon$, the term of order
$\epsilon$ integrates to zero in the dispersion relation for $S_\rho$, and the
corrections to the free-field approximation to $S$ are very small if
$\Lambda\le 3TeV$.  The leading order contribution to $T_\sigma$ is obtained by
neglecting the phase shift:
\begin{equation}
T_\sigma = {7\over 128\pi}\left({\Lambda\over M}\right)^4\epsilon_0, \;\;\;\;
\epsilon_0 = \left({\Lambda\over 4\pi v}\right)^2.
\end{equation}
$M$ is the parameter that determines the $WW\pi\pi$ form factor at $u=0$. If
$M\to\infty$ with $\Lambda$ fixed, we recover the one-loop result of the
nonlinear model:  $T_\sigma=0$.  If instead we identify $M = \Lambda$, we
obtain a contribution to the $\rho$ parameter
\begin{equation}
\delta\rho_\sigma = \alpha T_\sigma = \left
({M\over m}\right)^2{7g^4tan^2\theta\over
32768\pi^4} = 2.14\times10^{-4}
\left({M\over m}\right)^2{g^4tan^2\theta\over \pi^4},
\end{equation}
which agrees with the two-loop contribution~\cite{bij} in the standard model if
$M=1.65 m_H$, {\it i.e.}, for an effective cut-off of the order of $m_H$.  The
two-loop discrepancy between the nonlinear model (with $M\to\infty$)
and the large $m_H$ limit of
the standard model was emphasized by Bij and Veltman~\cite{bij}; this
discrepancy is removed when the appropriate higher dimensional operators are
included~\cite{appel},~\cite{long}.

If $s - M_i^2(s)$ vanishes for $m^2<s < \Lambda^2$, the phase shift goes
through
$90^0$ at $s=M_i^2$, and $\lambda_i(M_i^2)/M_i = \Gamma_i$ is the resonance
width.  Assuming resonance dominance, $M_i^2(s) \approx M_i^2\approx$ constant,
$\lambda_i =\Gamma_i M_i\approx$
constant, we can write the phase shifts (12) in the form
$$\delta_i(s) = {1\over 2i}\int_{-\infty}^sdt\left({1\over t - M_i^2
-i\lambda_i} - {1\over t - M_i^2 + i\lambda_i}\right) $$
\begin{equation}
 = \int_{-\infty}^sdt{\lambda_i\over (t - M_i^2)^2 + \lambda_i^2} .
\end{equation}
In the narrow width approximation: $\lambda_i\to 0$:
\begin{equation}
\delta_i(s) \to \int_m^sdt\pi\delta(t-M_i^2) = \pi\theta(s-M_i^2),
\end{equation}
and integration of the exponent gives the standard pole-dominated form factor:
\begin{equation}
|f_i(s)| = \left|{M_i^2\over M_i^2-s}\right|.
\end{equation}
To make the final integral finite we have to restore the finite width
in the expression for $|f|^2$ and then take the limit $\lambda_i\to 0$:
\begin{equation}
|f_i(s)|^2 = \lambda_i^{-1}{M_i^4\over (M_i^2 -s)^2 + \lambda_i^2}
\to \pi{M_i^3\over \Gamma_i}\delta(s-M_i^2).
\end{equation}
Using this result in (19) gives
\begin{equation}
T_\sigma = {9M_\sigma^7\over 512\pi^2v^2M^4\Gamma_\sigma}
[\ln(\Lambda^2/M_\sigma)^2 + 1].
\end{equation}
If we identify the scalar resonance with the Higgs particle of the standard
model, $M_\sigma = M= m_H$ and $\Gamma_\sigma = \Gamma_H = 3m_H^3/32\pi v^2$,
so we
recover the result in (1) (where only the divergent term was retained).  This
now appears as an order $g^2$ contribution to $\alpha T$ due to the factor
$m_H/\Gamma_H.$

For $J=1$ resonance dominance we get
\begin{equation}
S_\rho = {1\over 12\pi}\int_{m^2}^{\Lambda^2}{dt\over t}\pi{M^3_\rho\over
\Gamma_\rho}\delta(t-M_\rho^2) = {M_\rho\over 12\Gamma_\rho},
\;\;\;\; \Gamma_\rho = {g_\rho^2M_\rho\over 48\pi},
\end{equation} where $g_\rho$ is the $\pi\pi$-resonnance coupling constant on
mass shell.  The low energy prediction $f_\rho(0) = 1$ relates this to the
resonance-$W$ coupling $(g/2)\gamma_\rho$ [see Eq.(18)]: $g_\rho\gamma_\rho
= M^2_\rho$, so we recover the standard pole dominance result
\begin{equation}
S_a = {16\pi\over g^2}(\Pi'_{WW})_\rho
= {16\pi\over g^2}{\gamma_\rho^2\over M_\rho^4}.
\end{equation}
The pole dominance result for the axial $J=1$ channel is similar:
$$S_\pi = -{1\over 256\pi^3}\int_{m^2}^{\Lambda^2}{dt\over t}\pi{M^3_a\over
\Gamma_a}\delta(t-M_a^2) = -{M_a\over 256\pi^2\Gamma_a}$$
\begin{equation}
= -{16\pi\over g^2}(\Pi'_{WW})_a = -{16\pi\over g^2}{\gamma_a^2\over M_a^4}.
\end{equation} where $g_a$ is the $3\pi$-resonnance coupling constant on
mass shell and the resonance-$W$ coupling is $(g/2)\gamma_a$, $g_a\gamma_a =
M^2_a$; $\Gamma_a = g_a^2M_a/1024\pi^3.$  Such contributions have been
considered previously in the context of technicolor~\cite{stu},~\cite{maria}.

Present data, which constrain $|A^i|$ to be less than about 1~\cite{lang},
are insensitive to
the contributions given here if plausible values for the parameters are
assumed.
Better precision and/or a determination of the top quark mass might allow some
inference on strong $W,Z$ scattering cross sections that could be measured at
the SSC~\cite{chan}.  For example, present data mildly favor~\cite{lang} $S<0$;
interpreting this result without additional exotic contributions would lead
to the (counter-intuitive) conclusion of a larger average phase shift in the
axial channel than in the vector channel for $J=I=1$ (or $m_a<m_\rho$).
An important lesson is that for $m_H \sim TeV$, it is
likely to be incorrect to interpret the Higgs-sector contribution to $S$ and
$T$
in terms of a single parameter $m_H$.

\noindent{\bf Acknowledgement.}
\vskip 12pt
This work  was  supported  in  part by the
Director, Office of Energy Research, Office of High Energy and Nuclear Physics,
Division of High Energy Physics of the U.S. Department of Energy under Contract
DE-AC03-76SF00098 and in  part by the National  Science Foundation under grant
PHY-90-21139.


\vskip 28pt

\begin{thebibliography}{11}
\bibitem{tini} M. Veltman, {\it Acta Phys. Pol.} {\bf B8:} 475 (1977).
\bibitem{appel} T. Appelquist and C. Bernard, {\it Phys. Rev.} {\bf D22:} 200
(1980) and {\bf D23} 425 (1981).
\bibitem{chan} M.S. Chanowitz and M.K. Gaillard, {\it Nucl. Phys.} {\bf B261:}
379
\bibitem{stu} M. Peskin and T. Takeuchi, {\it Phys. Rev. Lett.} {\bf 65:} 964
(1990); M. Golden and L. Randall, {\it Nucl. Phys.} {\bf B361:} 3 (1991).
\bibitem{oren} O. Cheyette, {\it Nucl. Phys.} {\bf B297:} 183 (1988)
\bibitem{long} A. C. Longhitano, {\it Phys. Rev.} {\bf D22:} 1166 (1980),
{\it Nucl. Phys.} {\bf B231:} 205 (1994).
\bibitem{bij} J. van der Bij and M. Veltman, {\it Nucl. Phys.} {\bf B231:} 205
(1984).
\bibitem{omnes} Muskhelishvili, {\it Trud. Tbil. Mat. Inst.} {\bf 10:} 1
(1941).
R. Omn\`es, {\it Nuovo Cimento} {\bf 8:} 316 (1958);  M. Gourdin and A. Martin,
{\it Nuovo Cimento} {\bf 8:} 699 (1958).
\bibitem{mko} O. Cheyette and M.K. Gaillard {\it Phys. Lett} {\bf 197B:} 205
(1987); S. Dawson and S. Willenbrock, {\it Phys. Rev.} {\bf D40:} 2880 (1989);
M. Veltman and F. Yndurain, {\it Nucl. Phys.} {\bf B325:} 1 (1989);
R.S. Willey, {\it Phys. Rev.} {\bf D44:} 3646 (1991).
\bibitem{maria} D. Espriu and  M.J. Herrero, {\it Nucl. Phys.} {B373:} 117
(1992).
\bibitem{lang} T. Takeuchi, SLAC-PUB-5619, Lecture at the 1991 Nagoya Spring
School on Dynamical Symmetry Breaking, Nakatsugawa, Gifu-ken, Japan, April
23--27, 1991; P. Langacker, UPR-0492T,
Invited talk given at Int. Workshop on Electroweak Physics Beyond the
Standard Model, Valencia, Spain, Oct 2--5, 1991.
\end{thebibliography}
\end{document}